# Expanding Earth and Static Universe: Two Papers of 1935

## Helge Kragh[*]

**Abstract:** The German-born astronomer Jacob K. E. Halm (1866-1944) wrote in 1935 two papers on quite different subjects, one an astrophysically based argument for the expanding Earth and the other a no less original attempt to explain the galactic redshifts on the basis of a static universe. Of course, Halm was wrong in both cases. The second of the papers is reproduced *in toto* and compared to other early attempts to avoid the expansion of the universe by means of "tired light" explanations of the redshifts. Although often referred to in the literature on the expanding Earth, the content of Halm's first paper is not well known. This article also provides a brief account of Halm's life and scientific career, which included important studies of the solar spectrum (the "limb effect") and the first version of the mass-luminosity relation for stars.

## Introduction

Although a relatively minor figure in the history of astronomy, the German-British-South African astronomer Jacob Halm did interesting work in astronomy and its allied sciences [Spencer Jones 1945; Glass 2014]. This paper offers a brief account of Halm's career and some of his most important contributions to astronomy from the first two decades of the twentieth century. However, it focuses on two papers he published in 1935, nearly a decade after his retirement from his post at the Royal Observatory, Cape of Good Hope. The two papers, published in the same issue of the *Journal of the Astronomical Society of South Africa*, are not easily accessible and very rarely referred to in the history of science literature. Yet they are both of considerable interest, if for quite different reasons.

The first of the papers, a lengthy article based on Halm's 1934 presidential address to the Astronomical Society of South Africa, is a remarkable argument for the expansion of the Earth. Halm (1935a) justified the hypothesis primarily from astrophysical rather than geological reasons. While convinced that the Earth was in a state of expansion, he denied that the same was the case with the universe. In the second of the papers, Halm (1935b) developed a classical wave theory to explain the galactic redshifts on the basis of a static universe. He thus joined the "tired light" opposition against the new expanding-universe cosmology. A transcript of Halm's anti-expansion paper is included below. To put it into the proper historical context, I also briefly review other attempts from the period 1929-1937 to account for the Slipher-Hubble-Humason redshifts without assuming the expansion of space associated with relativistic cosmology.

[*] Niels Bohr Institute, Copenhagen University, 2100 Copenhagen, Denmark. E-mail: helge.kragh@nbi.ku.dk.

**Life and work of an astronomer**

Jacob Karl Ernst Halm was born 1866 in the German town Bingen am Rhein. After having received his doctorate at the University of Kiel on a mathematical subject, he was appointed assistant at the observatory in Strasbourg, at the time part of Germany (and spelled Strassburg). Halm stayed in Strasbourg 1889-1895 after which he moved to Scotland to take up a position at the Royal Observatory in Edinburgh. During his Edinburgh period he mostly worked with stellar spectra and spectroscopic studies of the Sun's rotation. In 1901 he became a naturalised British citizen and five years later he was elected a fellow of the Royal Astronomical Society. He resigned the fellowship in 1940.

On the recommendation of David Gill and Frank Dyson, in 1907 Halm was appointed chief assistant at the Cape Observatory in South Africa. In addition to his scientific work, Halm engaged in the organization of South African astronomy and especially in the Astronomical Society of South Africa of which he was a co-founder and served as president in 1924 and 1934. He was also active in the Cape Astronomical Association founded in 1912, ten years before the Astronomical Society [Coning 2012]. Halm retired in 1927 but remained active after he settled in Stellenbosch, where he gave lectures on advanced astronomy at the city's university founded in 1918.

While still in Edinburgh, Halm (1907) published a study in which he compared spectral lines of iron from the Sun's limb with those from its centre. He reported a small but systematic redshift near the limb which could not be explained as a result of the Sun's rotation or as a pressure effect. The "limb effect" – or what was sometimes called the "Halm effect" – was soon confirmed by leading astronomers, including George Ellery Hale and Walter Adams at the Mount Wilson Observatory. Halm and other researchers were unable to come up with a satisfactory explanation of the effect, which attracted further interest as a possible case of the gravitational redshift predicted by Einstein in 1911 [Forbes 1961; Crelinsten 2006, pp. 66-76].

Halm (1911) also studied the systematic motions of stars, finding evidence for a third class of "star streams" in addition to the two discovered by Jacobus C. Kapteyn. In this connection he argued that "the principle of equipartition of energy is applicable to the system of bodies constituting the visible universe." According to Halm, the average mass for a type of stars was inversely proportional to the square of the average velocity. Although Arthur Eddington (1914, pp. 159-160) found Halm's arguments interesting, he objected that they relied on a misguided analogy between the behaviour of stars and the molecules of a gas. Halm and Eddington met in August 1914, when the latter, together with other British astronomers, visited the Cape Observatory (see photograph in Glass 2014).

In the same 1911 paper Halm was the first astronomer to suggest a connection between the mass of stars and their luminosity and evolutionary state. He formulated this first version of the mass-luminosity relation as "intrinsic brightness and mass are in direct relationship." The relationship between mass and luminosity was further examined by Henry Norriss Russell and Ejnar Hertzsprung, and in Eddington (1924) later gave the full and theoretically argued version of it [Fernie 1969]. Based upon his studies of the distribution of stars, Halm (1917) concluded that light was absorbed along the galactic equator at a maximum amount corresponding to 2.1 mag kpc$^{-1}$. At the time most astronomers specializing in the Milky Way universe thought that interstellar absorption was negligible and probably less than 0.1 mag kpc$^{-1}$. Robert Trumpler's authoritative 1930 value for the absorption coefficient was 0.67 mag kpc$^{-1}$ [Seeley and Berendzen 1972].

**An astronomical approach to the dynamical Earth**

Halm's 1934 presidential address to the South African Astronomical Association was unusual by dealing principally with the Earth and not a more traditional astronomical subject.[1] On the other hand, his approach to the Earth was distinctly

---

[1] There is a striking parallel to Fred Hoyle's 1972 presidential address to the Royal Astronomical Society, where he used a new cosmological theory based on a decreasing gravitational constant to argue that the Earth is expanding at a rate $dR/dt \sim 0.1$ mm yr$^{-1}$. Neither Halm (1935a) nor Hoyle (1972) was trained in geology or geophysics and yet they dealt confidently with subjects belonging to the earth sciences. On Hoyle's address and it

astrophysical and entirely different from the traditional geological approach, which he criticized for being too limited and based on the axiom of a slowly contracting Earth. Halm insisted that the evolution of the Earth could only be understood on the basis of astrophysical theory and that such a perspective inevitably led to a very different picture, namely that the Earth had expanded through its entire history. Apparently unaware of earlier arguments for the expanding Earth [Carey 1988], he thought that his new picture was original.

From thermodynamic considerations of stellar and planetary atmospheres Halm obtained an equation which gave an invariant relation between the mean absolute temperature $T_s$ of a planet's (or star's) surface and its mean density $\rho$ measured in the unit g cm$^{-3}$:

$$\frac{T_s}{\rho^{8/21}} = \text{constant } (C)$$

We have found, he wrote, that

> … for every star there comes a moment when its life as an active gaseous body comes to an end. For reasons, the meaning of which we have not yet grasped, it is turned abruptly into a *rigid* body, the *rigor mortis* of star life has set in. … Once this rigor mortis has set in, the further fate of these star corpses is clearly defined by the condition in [the above] equation. The star *expands*, and the ratio between cooling and expansion is strictly regulated in accordance with [this] equation.

Using available data for stellar and planetary surface temperatures, he noted that white dwarfs and planets were characterized by approximately the same value, namely $C = 145$. The "fundamental equation," or "equation of evolution" as he also called it, could thus be written

$$T_s = 145 \times \rho^{0.38}$$

Halm emphasized that the equation was valid for all celestial bodies ranging from white dwarfs to the coolest planets. From the temperature-density law and certain speculative assumptions concerning the size of atoms at very high pressure he

---

derived that "at the beginning of geological time"[2] the radius of the Earth was 5430 km and its mean surface temperature about 63 °C. The radius $R$ of the primitive Earth would thus be less than the present one by 941 km or "about 100 times the height of Mount Everest." As to the average rate of expansion he estimated it to be $dR/dt \cong 1.6$ mm yr$^{-1}$ or "about the thickness of a penny-piece."

Since Halm assumed the Earth's mass to remain constant, in the geological past the density and surface gravity of the Earth would have been considerably higher than the present values. He estimated the original density to 9.13 g cm$^{-3}$ or 3.46 g cm$^{-3}$ larger than today. As to the surface temperature at two different epochs at which the radius of the Earth was $R_1$ and $R_2$, respectively, he calculated

$$\frac{T_1}{T_2} = \left(\frac{R_2}{R_1}\right)^{1.14}$$

For the past climate it meant that an increase in radius of one per cent corresponded to a 3.7 °C lower temperature. Contrary to later expansionists Halm argued that the primitive Earth was entirely covered by water, the continents only arising along with the expansion of the Earth.

Halm's theory of the Earth as a slowly cooling and expanding rigid body offered a new perspective on "the remarkable and fascinating suggestion regarding the formation of the continents made recently by the German geologist Wegener." At the time Alfred Wegener's theory of continental drift was not highly regarded and rejected by a majority of geologists and palaeontologists. The major supporter of drift in the 1930s was the South African geologist Alexander du Toit.[3] Halm's theory evidently differed from Wegener's, which assumed horizontal plate displacements on an Earth of constant size. While Halm admitted that there was no adequate physical force to move the continents in accordance with the drift theory, he nonetheless thought that Wegener's basic idea "is so strongly supported by their [the continents'] present configurations that it cannot be lightly rejected."

---

[2] Halm did not specify what this time was except that he took the separation of the continents to have begun approximately $10^9$ yr ago. In the 1930s the age of the Earth was typically assumed to be 2-3 × $10^9$ yr$^{-1}$. The presently accepted age of 4.5 × $10^9$ yr$^{-1}$ was only established in the mid-1950s.

[3] Halm probably knew du Toit, but I am not aware of any evidence that the two scientists were in contact.

In agreement with Wegener, Halm believed that the continents derived from a common supercontinent (Pangaea), only had this original continent split as a result of the expansion. He did not think of the expanding Earth as an alternative to continental drift, but rather as an improved version of it. "The single conception of the Earth as an *expanding* body," he wrote, "has based Wegener's fascinating theory on a sound physical principle and has opened new vistas of approach towards the solution of the many problems which the history of our planet lays before us."

Halm's theory attracted very little attention, but some later proponents of expansionism considered it a precursor of what in the 1960s emerged as the modern expansion theory of the Earth [Carey 1988, p. 140]. Although the expansion of the universe was well known among astronomers by 1935, Halm did not suggest a connection between planetary expansion and cosmic expansion. Indeed, he denied that the universe expanded. All the same, a few later expansionists did suggest a connection [MacDougall et al. 1963]. With Hubble's law applied to the Earth

$$v = H_0 r \, , \quad r = R_E = 6370 \text{ km} \, ,$$

the expansion rate becomes $v = 0.66$ mm yr$^{-1}$ if based on $H_0 = 100$ km s$^{-1}$ Mpc$^{-1}$, a value widely accepted in the 1960s. According to some leading geophysicists, the Earth expanded at the same rate. Could this be just a coincidence? In the mid-1930s the accepted value of Hubble's constant was about 5 times larger and would thus have resulted in $v \sim 3$ cm yr$^{-1}$ or considerably larger than the expansion rate assumed by Halm. But this is of course just a side remark.

**Early alternatives to the expanding universe**

Although Halm's 1935 paper on non-recessional galactic redshifts contained no references, he can hardly have been unaware that his alternative to the expanding universe was not the first of its kind. Many scientists (and most non-scientists) found it difficult to accept the concept of an expanding cosmic space, and no less difficult to accept that the distant galaxies recede from us at the furious speeds of $v = 4 \times 10^4$ km s$^{-1}$ or more. They consequently came up with alternatives to explain the redshift data, which provided the sole empirical evidence for the expanding universe. The table below gives a list, undoubtedly incomplete, of alternatives proposed in the period 1929-1937.

| Author | Background | Year | Comment |
|--------|-----------|------|---------|
| F. Zwicky | astronomy | 1929 | gravitational "drag" mechanism |
| J. Stewart | astrophysics | 1931 | no specific mechanism |
| W. MacMillan | astronomy | 1932 | no specific mechanism |
| H. Buc | ? | 1932 | speculation on loss of light energy |
| W. Nernst | physical chemistry | 1935 | $H_0$ interpreted as "quantum constant" |
| J. and B. Chalmers | physics | 1935 | assumes increase of $h = h(t)$ |
| P. Wold | physics | 1935 | assumes $c = c(t)$ |
| J. Halm | astronomy | 1935 | classical wave explanation |
| S. Sambursky | physics | 1937 | assumes $h$ to decrease with time |

While many astronomers and physicists had accepted the expanding universe by the mid-1930s, others resisted it and still more preferred an agnostic attitude. As a representative for the last group may be taken the Canadian astronomer John Stanley Plaskett, director of the Dominion Astrophysical Observatory in Victoria, British Columbia. In an address on the current situation in cosmology, Plaskett (1933, p. 243) stated:

> Either this increasing red shift is a Doppler effect and the Universe … is expanding at an alarming rate, or else the displacement is due to some action on the light producing atoms which decreases their vibration frequency, hence increasing the wave length and shifting the lines to the red, proportionally to the distance of the source. It must be remembered that there is no known means of determining from the nebular spectra which of these two processes is operative.

Remarkably, Edwin Hubble largely shared the cautious agnosticism expressed by Plaskett. In his influential *The Realm of the Universe*, Hubble (1936, p. 122) wrote, "judgement may be suspended until it is known from observations whether or not red-shifts do actually represent motion."

The first and most influential alternative to galactic recession was proposed as early as August 1929 by Fritz Zwicky at the California Institute of Technology. Zwicky was the founder of "tired light" mechanisms, a term that in general refers to the idea that photons slowly lose energy on their journey through space and therefore are observed with an increased wavelength. The name "tired light" is often ascribed to Richard Tolman (always without a proper reference), but it seems to have been coined by the Princeton physicist Howard Percy Robertson (1932), who in an address of 29 April 1932 referred to the hypothesis that "the observed red shift would be due to the properties of 'tired' light rather than the nebulae themselves." Robertson, a leading relativist cosmologist, found explanations of this kind to be unsatisfactory and *ad hoc*. The name may have been used informally at earlier occasions, as Stewart (1931) referred to "what has been called the 'fatigue' of light quanta."

Zwicky (1929) discussed a number of possible explanations of the spectral shifts, including that they might be due to an ordinary photon-electron Compton effect or a galactic gravitational field. These explanations he dismissed as inadequate and instead focused on what he called a "gravitational analogue of the Compton effect." According to this mechanism, light of frequency $\nu$ travelling a distance $r$ would lose energy corresponding to the redshift

$$\frac{\Delta \nu}{\nu} = \frac{1.4 G \rho D}{c^2} r$$

The quantity $D \gg r$ is a measure of the distance over which the gravitational "drag" operates, and $\rho$ is the average density of matter in the universe, which Zwicky took to lie in the interval $10^{-25} > \rho > 10^{-31}$ g cm$^{-3}$. He found the gravitational-drag explanation to be "in qualitative accordance with all of the observational facts known so far." As support of his theory he referred to discussions with the young German astronomer Paul ten Bruggencate, who at the time stayed at the Mount Wilson Observatory. Inspired by Zwicky, Bruggencate (1930) studied the radial velocities of globular clusters, obtaining results of the order $\Delta \nu / \nu \sim 10^{-3}$ that "reconcile the observed red-shift with Zwicky's prediction."

In a couple of later papers Zwicky (1933; 1935) returned to his theory characterized by a redshift that not only depended on the distance but also on the distribution and amount of cosmic matter. The first of these papers is today best known for its bold prediction of dark matter. Zwicky (1935) admitted that the gravitational-drag theory was strongly hypothetical and not entirely satisfactory. On the other hand, it had the methodological advantage that it was empirically testable: "An initially parallel beam of light, on this theory, will gradually open itself because of small angle scattering. Observational tests on this point will be important."

Most tired-light hypotheses in the 1930s were unsophisticated compared to Zwicky's original hypothesis. In many cases they were nothing but guesswork. John Quincy Stewart (1931), an astrophysicist at Princeton University, looked for a numerical connection between Hubble's recession constant and other constants of nature. With $m_0$ denoting the electron's rest mass he came up with

$$H \cong \frac{e^6}{hGm_0^3 c^3} = 1.37 \times 10^{27} \text{ cm},$$

which he compared to the galactic recession value found by Hubble and Humason, $H_0 = 538$ km s$^{-1}$ Mpc$^{-1}$ = $1.66 \times 10^{27}$ cm. Stewart suggested that photons lost their energy ($E = h\nu$) in proportion to the distance $x$ from the source according to

$$\nu(x) = \nu(0)\exp\left(-\frac{x}{H_0}\right)$$

He did not offer a physical explanation for this simplest form of tired-light hypothesis.

A tired-light explanation very similar to Stewart's was offered by H. E. Buc (1932) to avoid the "most amazing speculation" of an expanding universe, and also, the same year, by the respected Chicago astronomer William Duncan MacMillan. For long an advocate of an eternal, stationary and self-perpetuating universe with matter and energy in steady interaction, MacMillan had no taste for relativistic cosmology. He speculated that the energy evaporated by the galactic photons "disappears into the fine structure of space and reappears eventually in the structure of the atom" [MacMillan 1932; Kragh 1995]. As another possibility he mentioned that the evaporated energy might still exist as a kind of low-frequency cosmic background radiation. However, writing in 1932 and not after 1965, "there is at present no evidence of such radiation."

MacMillan's conception of the universe was to a large extent shared by the famous physical chemist Walther Nernst, who in the 1930s occupied himself more with astronomy and cosmology than with chemistry [Kragh 1995]. No wonder that he resisted the expanding universe and suggested a redshift explanation similar to the ones of Stewart and MacMillan. Nernst (1935; 1937) reasoned that from the assumption $dE/dt = -HE$ it follows that

$$\ln\left(\frac{v_0}{v}\right) = Ht$$

Since $t = r/c$ and the decrease in frequency satisfies $\Delta v \ll v$, one gets

$$\frac{\Delta v}{v} = \frac{H}{c}r \,,$$

which is the empirically confirmed Hubble law. As Nernst saw it, the constant $H$ was not really a constant of the universe, but a "quantum constant" giving the decay rate of photons. Noting that $hH \sim 10^{-71}$ J has the dimension of energy, he speculated that the quantity might be a minimum energy characterizing the zero-point energy of the universe which he had earlier hypothesized [Kragh and Overduin 2014, pp. 29-38]. Moreover, Nernst calculated that if the loss in photon energy was absorbed in the ether filling intergalactic space, it would provide it with a constant background temperature of $T = 0.75$ K.

Variation in a photon's frequency in free space does not necessarily conflict with energy conservation. Assuming $E = hv = $ constant implies

$$v\frac{dh}{dt} + h\frac{dv}{dt} = 0$$

What if Planck's constant is not the same at the time a photon leaves a galaxy and at the time it is received on Earth? Hypotheses of $h = h(t)$ were proposed by a few physicists in the 1930s. J. A. Chalmers and Bruce Chalmers (1935a; 1935b) suggested a non-Doppler interpretation by assuming $h$ to increase exponentially with a doubling time of about $1.4 \times 10^9$ yr. "It has been shown experimentally," they wrote, "that $h$ is a constant here now, but there is nothing except the Hubble effect that can possibly give information on the value of $h$ in the remote past. We are thus interpreting the Hubble effect as an alteration of $h$ as between emission in the nebula and emission here now." The two British physicists thought that their hypothesis might help overcoming the time-scale difficulty common to most expanding models of the universe.

In part inspired by Eddington's attempt to unify cosmology and quantum physics, there was in the period much interest in the constants of nature. While Chalmers and Chalmers assumed $h$ to increase, Samuel Sambursky at the Hebrew

University in Jerusalem suggested reconciling the static universe and the observed redshifts by assuming $h$ to *decrease* with time. Sambursky (1937) assumed

$$h(t) = h_0 \exp(-Ht)$$

and identified $H$ with the Hubble factor given by

$$H = -\frac{1}{h}\frac{dh}{dt}$$

With $H$ = 485 km s$^{-1}$ Mpc$^{-1}$ he thus found $dh/dt \cong -10^{-50}$ J for the variation of $h$.

Finally, if $h$ is allowed to vary with time, why not other fundamental constants, such as the elementary charge $e$ and the velocity of light $c$? Peter Wold (1935), a physicist at Union College, Schenectady, assumed that $c$ varied as

$$c = c_0(1 - kt)$$

With an appropriate value for $k$, namely $k$ = 5.7 × 10$^{-10}$ yr$^{-1}$, this "simple assumption leads one definitely to a redward shift … in agreement with the Hubble-Humason observations." Unfortunately it also led to the consequence that, if the rate of decrease continued, $c$ = 0 in less than 2 billion years! For this reason Wold added another *ad hoc* hypothesis, that $c$ varied periodically with a very long period of time. The idea of a varying speed of light was entertained by a few other researchers in the 1930s and much later it would enter modern cosmology [Kragh 2006].

The mentioned non-expansion redshift hypotheses, whether belonging to the tired-light category or not, were only the beginning of a minor industry that has continued to this date. It only started accelerating in the 1950s with a paper by Erwin Finlay-Freundlich (1954), a former collaborator of Einstein. However, it is not the purpose of this paper to review or evaluate the whole class of tired-light hypotheses, but only to offer a comparative perspective on Halm's 1935 paper.

**A wave theory of galactic redshifts**

Halm (1935b) considered the relativistic explanation of the expanding universe to be a "maze of abstruse speculations," whereas his alternative was simple and "self-evident." It is generally assumed, he says, that if the galactic source does not move with respect to the Earth, and light traverses in free space, the wavelength (or frequency) of a monochromatic ray of light remains unaltered. However, he questions this assumption and purports to show that a wave may undergo adiabatic expansion (or contraction), meaning that the potential energy increases steadily. By assuming the wave expansion to vary linearly with time he arrives at a frequency displacement proportional to the distance.

Contrary to the advocates of tired-light hypotheses Halm does not appeal to interaction between light and matter, or between light and gravitational fields. There also is no "evaporation" of light energy in his proposal, which consequently does not belong to the tired-light category. Halm does not refer to light as consisting of quanta or photons satisfying $E = h\nu = hc/\lambda$. Indeed, quantum theory and Planck's constant do not appear in his scheme, which is based solely on a classical analysis of wave motion.

Below follows an exact transcript of Halm's paper:

---

## On the Theory of an "Expanding Universe"

By J. K. E. Halm, Ph.D., F.R.A.S.

The conception of an *expanding universe* is based on the observation that certain lines in the spectrum of distant clusters are displaced towards the less refrangible side and that this displacement increases in proportion to the distance of the cluster. Since we are accustomed to interpret displacements of spectral lines by motion in the line of sight, the inference has been drawn that the Universe *in toto* must recede from us. The essential condition on which alone this conclusion could be warranted is that, in absence of motion in the line of sight, the wavelength of a monochromatic ray remains unaltered whatever the distance between source and receiver may be. Before accepting the reality of the enormous velocities with which we shall have to endow the distant members of the Cosmos, if the observed displacements are really due to motions, we are justified in demanding a convincing proof of the correctness of the assumption that the wavelength of the ray remains unaltered however long its journey through space may be.

From the most general dynamical aspect a wave represents a system in which the motion repeats itself in all respects at certain intervals of time which may be denoted by $i$. If $T$ denotes the kinetic energy at any moment, $V$ its potential energy, and if $T_m$ and $V_m$ signify the average values of these quantities taken over the interval of recurrence, $i$, any small disturbance affecting the system will produce changes $\delta T_m$, $\delta V_m$ and $\delta i$ which must satisfy Hamilton's fundamental equation. In a recurrent system this equation has the form

$$2\delta(T_m i) = i\delta(T_m + V_m) = i\delta E \ ,$$

or

$$\delta T_m - \delta V_m = -2T_m \frac{\delta i}{i} \tag{1}$$

$E$ representing the total energy. (See Routh, Advanced Rigid Dynamics, 4th Edition, Art. 461.)

In addition, the law of the conservation of energy supplies the equation:

$$\delta E = \delta(T_m + V_m) = 0 \tag{2}$$

Consequently the second equation (1) assumes the form:

$$\delta T_m = -\delta V_m = -T_m \frac{\delta i}{i} \tag{3}$$

Dynamics thus supplies two equations for three variables. A unique solution, therefore, is possible only when a third equation can be established. This equation must necessarily be of a purely *empirical* character, *i.e.*, it must be framed in such a manner that it satisfies the results of observation. Thus, while equation (2) establishes the constancy of the *sum of the two energies*, $(T_m + V_m)$ on the basis of a general dynamical principle, viz. the conservation of energy, we are not permitted *a priori* to conclude that these energies are individually constant, *i.e.*, $\delta T_m = 0$ *and* $\delta V = 0$. Theoretically, the energies of a wave on its journey through space may be subject to changes $\delta T_m = 0$ *and* $\delta V = 0$, provided that equation (2) is always satisfied, *i.e.*, the wave may undergo *adiabatic* transformations. It may expand or contract or remain stationary. Its *observed* behaviour alone can decide on these alternative possibilities.

The most general form in which the third equation can be framed is to consider $T_m$ as some function of the time $t$ and write:

$$T_m = f(t); \quad \frac{\delta T_m}{\delta t} = \frac{\delta f}{\delta t} \ ; \ \frac{\delta V_m}{\delta t} = -\frac{\delta T_m}{\delta t} = -\frac{\delta f}{\delta t} \tag{4}$$

But from equation (3) we find by integration:

$$T_m i = h \, , \tag{5}$$

where $h$ is a constant. Considering that $i$ has been defined as the time interval of one complete recurrence, $1/i$ represents the number of recurrences in unit time, *i.e.*, the frequency of the oscillations which is usually denoted by $\nu$. Hence

$$\frac{\delta T_m}{\delta t} = h \frac{\delta \nu}{\delta t} = \frac{\delta f}{\delta t} \tag{6}$$

Denoting by $\nu_0$ and $f_0$ the values at the moment when the ray leaves the cluster and by $\nu$ and $f$ the values at the time of arrival in the spectroscope we write

$$\nu - \nu_0 = \frac{1}{h}[f - f_0] \tag{7}$$

The equation refers to a state of relative rest in the line of sight. If the cluster moves with a radial velocity, $v$, the difference in the frequency is expressed by

$$\nu - \nu_0 = \frac{1}{h}[f - f_0] - \frac{\nu_0}{c} v \tag{8}$$

where $c$ represents the velocity of light.

Let us now assume that the potential energy increases progressively, *i.e.*, that the wave *expands adiabatically*, and that this expansion is proportional to the time. Obviously in this case

$$f - f_0 = -\alpha[t - t_0] \, ,$$

where $\alpha$ is supposed to be an extremely small quantity. Since the distance $r$ between cluster and star is:

$$r = c(t - t_0),$$

$$f - f_0 = -\frac{\alpha}{c} r \, ,$$

and equation (8) becomes:

$$\nu - \nu_0 = -\frac{\alpha}{ch} r - \frac{\nu_0}{c} v \tag{9}$$

The theory of the *Expansion of the Universe* represents the special case $\alpha = 0$, *i.e.*,

$$\nu - \nu_0 = -\frac{\nu_0}{c}v$$

The observed shift of the lines is attributed entirely to motions in the line of sight.

The theory of an *adiabatic expansion of the wave* attributes the observed progressive decrease in the frequency with the distance *r* to this expansion. The displacements due to velocity appear in the character of *accidental* errors, *i.e.*, plus and minus alike.

As pointed out, from the dynamical point of view both assumptions are possible. The choice lies between a theory which so far has entangled the mind in a maze of abstruse speculations without offering a definite hope of solution, and, on the other hand, an assumption which leads directly to a self-evident explanation of the observed phenomena.

---

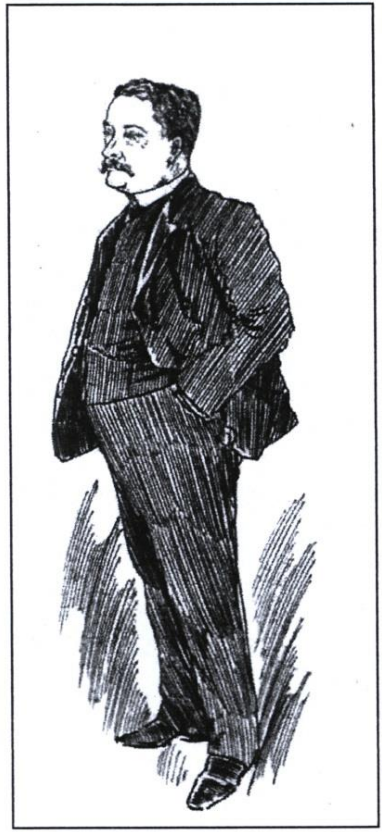

Cartoon of Halm, Cape Times, 19 September 1908. Reproduced from Glass 2014.